\documentclass[12pt, preprint]{aastex}
\usepackage{graphicx}
\shorttitle{An observational revisit of band-split solar type-II radio bursts} \shortauthors{Du et al.}
\usepackage{color}

\usepackage{lineno}
\usepackage{multirow}

\begin{document}

\title{An observational revisit of band-split solar type-II radio bursts}

\author{Guohui Du\altaffilmark{1},Xiangliang Kong\altaffilmark{1},Yao Chen\altaffilmark{1},
Shiwei Feng\altaffilmark{1},  Bing Wang\altaffilmark{1}, and Gang
Li\altaffilmark{2}}

\altaffiltext{1}{Shandong Provincial Key Laboratory of Optical
Astronomy and Solar-Terrestrial Environment, and Institute of
Space Sciences, Shandong University, Weihai 264209, China;
yaochen@sdu.edu.cn}
\altaffiltext{2}{Department of Space Science and CSPAR, University
of Alabama in Huntsville, Huntsville, AL 35899, USA}

\begin{abstract}
Band split of solar type II radio bursts, discovered several decades ago,
is a fascinating phenomenon with the type-II lanes exhibiting two almost-parallel
sub-bands with similar morphology. The underlying split mechanism remains elusive.
 One popular interpretation is that the splitting bands are emitted from the shock
 upstream and downstream, respectively, with their frequency ratio ($\gamma$)
 determined by the shock compression ratio. This interpretation has been taken
 as the physical basis for many published references. Here we report an
 observational analysis of type II events with nice split selected from the
 ground-based RSTN data from 2001 to 2014, in the metric-decametric wavelength.
 We investigate the temporal variation and distribution of $\gamma$, and
 conduct correlation analyses on the deduced spectral values. It is found
 that $\gamma$ varies in a very narrow range with $>$80\% of $\gamma$ (one-minute
 averaged data) being between 1.15 to 1.25. For some well-observed and long-lasting
 events, $\gamma$ does not show a systematic variation trend within observational
 uncertainties, from the onset to the termination of the splits. In addition,
 the parameters representing the propagation speed of the radio source (presumably
 the coronal shock) show a very weak or basically no correlation with $\gamma$.
 We suggest that these results do not favor the upstream-downstream scenario of
 band splits.
\end{abstract}


\section{Introduction}
Solar type II radio bursts are excited by energetic electrons
accelerated during coronal eruptions (\citealp{Payne1947,
Wild1963, Nelson1985}). It is generally believed that coronal
shocks are accelerators of these electrons (see,
\citealp{Chen2014, Feng2015, Kong2015}, for latest studies).
Despite intensive studies over half a century, it has not
been completely understood how these electrons are accelerated at
shocks, and how the radio emissions are excited.

One fascinating feature of solar type II radio bursts is the band
split \citep{McLean1967, Wild1972, Smerd1974}. This feature can
appear on both the fundamental and harmonic branches. The physical
mechanism underlying this phenomenon remains unresolved during the
past decades. Proposed scenarios have been briefly reviewed
 in \citet{Du2014}, and can be divided into two classes. In the
 first class it is assumed that the split is caused by the shock
 structure, such as the upstream, downstream, as well as the transition
 region across the shock \citep{Smerd1974, Smerd1975}, or the shock interaction
 with coronal structures at different locations characterized by
 different emission frequencies \citep{McLean1967}. In the other
 class of scenarios it is assumed that the split is associated
 with some intrinsic emission mechanism \citep[e.g.,][]{Treumann1992}.
  It should be pointed out that none of the available scenarios
  receives solid observational or theoretical supports, so the
  exact mechanism of band splits remains unknown. Nevertheless,
  in many published references, the assumption that
  the split bands are respectively emitted from downstream
  (the upper band with higher frequency) and the upstream (the
  lower band with lower frequency) has been used for shock
diagnostics, for example, to infer the compression ratio and the
shock Mach number, etc. (e.g., \citealp{Smerd1974, Vrsnak2002,
 Vrsnak2004, Cho2007, Liu2009, Ma2011, Vasanth2014, Zucca2014, Long2015}).
 However, the upstream-downstream (UD) scenario suffers
from the absence of energetic electrons and/or enhanced Langmuir
oscillations in the shock downstream, either theoretically or
observationally, (see \citealp{Cairns2011} for a recent review).

Thus, it is important to search for observational clues
 and physical mechanisms accounting for the band-split phenomenon.
In a series of studies, \citet{Vrsnak2002,Vrsnak2004}
investigated statistically the corona and interplanetary type-II
bursts with band split. Assuming the UD scenario, they further
deduced the magnitude and radial profile of the Alfv{\'e}n speed and
magnetic field strength. Differences of their studies and ours
will be discussed later in this paper. In this study we present
an observational revisit of band-split type II events by examining
the radio spectral data from the ground-based RSTN (Radio Solar
Telescope Network, operated by the US Air Force) in the metric to
decametric wavelength.

\section{Events selection and their general properties}\label{sec2}

The type-II band-split (BS) events are selected according to their
spectral characteristics, and we focus on the ground-based
observational data that are in the metric to decametric
wavelength. Following \citet{Du2014}, we require that the two
bands should be narrow and clear enough with similar intensity and
morphology variations, and without strong interference from other
type of solar bursts or artificial signals. In addition, the
frequency ratio of the upper and lower bands should be less than 2
to avoid confusion with the fundamental and harmonic branches
\citep[see also, ][]{Vrsnak2001}. We also require that the events
should last for more than one minute. We examine the RSTN data
from 2000 to 2014, and select 18 candidate events.

Mostly according to the lifetime and clearness of the band splits, we
further separate the events into two groups. Group A events have a
lifetime longer than 10 minutes with generally nice BS features. This
allows us to examine the temporal evolution of the split parameters
during an individual event. Events in group B either have a shorter
lifetime or their BS features are not as clear as those of group A
events.

The spectra of group A events are given in Figure 1, and
some basic parameters of all the events (referred to as Group A+B)
have been listed in Table 1 including the accompanying CME and
flare properties. For further quantitative analysis, we have
over-plotted the backbones of the splitting bands with dashed
lines. These backbones are given by the intensity maximum. Only
the better-observed branches are superposed with their backbones.
In further studies, we will measure the exact values of
these frequencies and calculate the average frequency drift and
the split ratio. The frequencies will be divided by 2 if
they are taken from the harmonic branch.

In Table 1, the first three columns show the event dates, the
start and end times of the type II emission, and the interval used
in this study which is selected to avoid data with unclear split
feature or strong interference. Further spectral parameters given
in the following columns are deduced within this selected
interval. The start and end frequencies of the lower band
($f_{start}$ and $f_{end}$), the average drift (Df/Dt), and the
average band-split frequency ratio ($\gamma$) are given in the
fourth to sixth columns. The first appearance time of the coronal
mass ejection (CME) in the LASCO-C2/SOHO \citep{Domigno1995} field
of view, and the flare data are given in the left columns. The CME
and flare data with superscripts are obtained by the authors while
others are taken from the CDAW-CME catalogue \citep{Yashiro2004}
and the SolarSoft Latest Events Archive
($\url{http://www.lmsal.com/solarsoft/latest\_events\_archive.html}$).
Note that in Events 20140220 and 20140907, two episodes of split
bands can be identified, and we list them separately in Table 1.

The most striking feature of Figure 1 is the similarity of the
frequency ratio of the two splitting bands for different events.
Note that these events are observed during different phase of the
solar cycle, with significant difference in the spectral drift and
CME-flare parameters. This can also be seen from Table 1 that the
average frequency split ratio varies from 1.14 to 1.29 while the
average drift varies in a large range from -0.03 to -0.23 MHz
s$^{-1}$. In addition, the CME linear speeds also vary
significantly from $\sim$ 200 km s$^{-1}$ to $\sim$ 1300 km
s$^{-1}$, and the flare levels vary from C1.1 to X1.0.

Comparing the timing data given in Table 1, we find that for all events,
the type II starting time is earlier than the CME first appearance time
in LASCO C2, and later than the flare start times. In addition, the type
II bursts start within 5 minutes of the flare peaking time for 12 events.
This is consistent with previous statistical correlation studies of the
type II bursts and solar eruptions \citep[e.g., ][]{Mancuso2004}. It is
not possible to infer the exact cause (i.e., CME or flare) of the radio
bursts only with these timing data.

To further illustrate the similarity of different BS events, in
Figure 2(a) we superpose the backbones of the eight events in
Group A. In doing so, we take the two-fold (2X) Saito density
model \citep{Saito1970} as the reference model. Data on each
backbone are moved in time as a whole until a minimum deviation
from the backbone to the 2X Saito model is achieved. We see that
all these temporally-shifted backbones, being put together, still
exhibit a recognizable BS morphology. This highlights the obvious
spectral similarity among different data sets.

In Figure 2(b), we plot the frequency ratios ($\gamma$) of the
splitting bands together with the same starting time. Two features
should be emphasized. Firstly, a majority of $\gamma$ lies in a
narrow range of 1.15 to 1.25 with only a few outliers. Secondly,
we find that within observational uncertainties $\gamma$ does not
vary much in time for individual events. To show the uncertainty
of the data, we determine the frequency profiles given by 90\% of
the intensity maximum for one specific event (20020125) which is
long enough to cover most part of the frequency range. This
produces four spectral lines. We then take the ratios of the outer
two lines and the inner two lines to be the maximum and the
minimum of the $\gamma$ uncertainty. Other events have similar
uncertainty range, and will not be shown here for clearness. These
two features, i.e., the values of $\gamma$ do not change
significantly from event to event and during the evolution of a
specific event, are consistent with previous studies on fewer and
earlier events (see, e.g., \citealp{Vrsnak2001}). We confirm this
finding with more events in this study.

To take into account the large spectral changes during an event,
we calculate the one-minute averaged values of $\gamma$. To do
this, the data are separated into one-minute long episodes with
adjacent ones having an overlap of 30 seconds. For an event
lasting for 5 minutes, this produces 9 data points of the averaged
$\gamma$. Histograms of the number of data points with the
one-minute averaged values of {$\gamma$} for Group A and Group A+B
are given in Figure 2(c) and Figure 2(d), respectively. We see
that for both data sets, the mean and median values of $\gamma$
are $\sim$1.2, and the standard deviation (SD) is 0.03 for
Group A and 0.04 for Group A+B. For Group A (A+B), $\sim$95
(80)\% of the data points lie in a narrow range of [1.15, 1.25].
This result is consistent with the conclusion obtained above.

\section{Correlation analysis of the splitting frequency ratio ($\gamma$) and other spectral parameters}\label{sec3}

It is generally believed that the type II radio bursts are given
by plasma emission with the emission frequency mainly determined
by the electron density at the radio source (Ginzberg \&
Zhelezniakov 1958). So the spectra data of type II bursts can be
used to deduce the source velocity using a density model of the
corona. Among various density models, the Saito \citep{Saito1970}
and Newkirk \citep{Newkirk1961} models have been widely used in
relevant studies. The 2X Saito model is very close to the Newkirk
model in the inner corona, and both models are represented with
multinomial function of the radial distance.

Similarly, we make a simplified assumption that the corona density
and the type-II frequency can be approximated by the following power
laws of the radial distance (see also \citealp{Gopalswamy2009}),
\begin{equation}
n=n_{0}r^{-{\alpha}},    f=f_{0}r^{-\frac{\alpha}{2}}
\end{equation}
where $n_0$ and $f_0$ are the number density and plasma frequency
at the coronal base. Assuming the radio source (presumably the
coronal shock) velocity is $v_{s}$, we have
\begin{equation}
\frac{Df}{Dt}=-\frac{v_{s}}{2} \alpha
{f_{0}}^{-\frac{2}{\alpha}}\cdot f^{\frac{\alpha+2}{\alpha}}
\end{equation}
\begin{equation}
v_{s}\propto - \frac{2}{\alpha} \frac{Df}{Dt} f^{-\frac{\alpha
+2}{\alpha}}= v_{s}^{p}
\end{equation}

From previous studies, it is well known that the type-II spectral
drift is strongly correlated with the emission frequency. This
indicates that lower in the corona the density gradient is
larger and thus the spectral drift is higher, in general
\citep[e.g., ][]{Vrsnak2001, Gopalswamy2009}. This relationship
has been confirmed by our data analysis shown in the upper panels
of Figure 3, where we present the one-minute averaged spectral
drift ($\frac{Df}{Dt}$) versus $f^{\frac{\alpha+2}{\alpha}}$ with
{$\alpha$} set to 2, 4, 6, and 10 in different panels for all
events. We also plotted the linear fitting results with black
lines. The obtained correlation coefficients $c$ for different
values of {$\alpha$} are around 0.67 and very close to each other,
consistent with previous studies.

From Equation (3) we can deduce a proxy of the radio source speed
({$v_{s}^{p}$}) with the spectral data if given the power law
index $\alpha$. We then check the correlation of this source speed
proxy with the band split frequency ratio {$\gamma$}. The results
are given in the lower panels of Figure 3 for different
assumptions of {$\alpha$}. In each panel we also present the
linear fitting line and the corresponding correlation coefficient
($c$). It is found that in panel (e) {$c$ = 0.2} for $\alpha = 2$,
while for other values of $\alpha$ $c$ is less than 0.1. This
indicates a general weak or even no correlation between the shock
speed proxy and the type-II split ratio.

In Figure 4, we plot the deduced radio-source speed
with the 2X Saito density model by fitting the radio spectra to
the lower band of the type II splits. Again, we separate the data
into episodes of one minute with adjacent episodes sharing 30
seconds of data. The correlations of the obtained source speeds
and the frequency split ratios of the events in Group A and Group
A+B are given in panels (a) and (b). The correlation coefficients
are only marginally larger than 0.1 indicating very weak
correlations. The weak or even no correlation described above is
another major result of this study.

\section{Conclusions and discussion}\label{sec4}
With the RSTN radio spectral data, we collected 18 type II
events with nice band-split features. The striking similarities
of the splitting bands and the relative constant frequency
ratio of the two splitted bands are highlighted. It is found
that a majority ($> 80 \%$) of the split frequency ratio data
lie in a very narrow range from 1.15 to 1.25, which does not
show a considerable systematic change during most individual
events, within measurement uncertainty. In addition, we find
that there exist very weak or basically no correlation between
the prescribed proxy of the source velocity and the frequency
ratio of the split bands. These results, based on the following
discussion, do not favor the mostly-used UD assumption of the
type-II band splits.

According to the UD scenario, the band-split ratio should strongly
depend on the shock properties. These properties are determined by
the coupling between the coronal eruption and the background
environment. During the progression of the solar cycle, the
coronal environment changes significantly, and the solar eruptions
also exhibit a large variation (see our Table 1). Combining these
two varying factors, it is expected that the shock properties vary
significantly from event to event. Furthermore, when the UD
scenario was first proposed by \citet{Smerd1974}, the shock was
considered to be an ideal discontinuity. In reality, in the shock
upstream and downstream, strong turbulence as well as the shock
foot and overshoot structures may develop
\citep{Li2013,Schwartz2011,Scudder1986}. This certainly adds more
variation effect to the upstream-downstream density ratio.
Therefore, if the UD scenario is correct, at least the following
three outcomes shall be expected.

Firstly, the density ratio shall present a large variation from
event to event. In addition, from the onset to the termination of
a specific type-II event, the shock may evolve significantly and
so likely does the density compression. These expectations are not
observed in our data analysis. Secondly, the split frequency
(presumably the density compression ratio according to the UD
scenario) should strongly depend on the shock properties. The
shock itself is considered to be related to a weak diffusive
structure in the coronagraph and possibly associated with the EUV
wave front in the EUV imaging data (see, e.g., \citealp{Chen2014,
Feng2015}), whose properties are still very difficult to measure.
So a direct examination of the shock compression measurement and
the band-split ratio is very difficult, if not impossible. In this
study, we presented a correlation analysis of the frequency split
ratio and a velocity proxy of the radio-emitting source. The proxy
is deduced using the prescribed power-law coronal density model
and the frequently-used Saito density model. No correlation or
only marginal correlation between the source velocity proxy and
the split ratio is found. This does not lend to a strong support
to the UD scenario.

Thirdly, with the UD scenario one is able to deduce the shock
compression ratio with the obtained frequency split ratio. The
deduced compression ratio is in general very small and lies in a
narrow range of [1.2, 1.7]. This suggests that the split
is highly selective over the compression ratio, corresponding to
relatively weak shocks. This result is not consistent with the
intuitive expectation that energetic electrons as well as the
type-II bursts are more likely associated with stronger shocks
\citep[e.g., ][]{Wu1984}.

Another evidence against the UD scenario was presented by
\citet{Du2014}, who found a BS event with spectral features
carried by the high-frequency band appearing seconds earlier than
those carried by the low-frequency band. According to Du et al.,
these spectral features were related to density structures which
were swept by the shock during its propagation. Assuming the UD
scenario, the density structures shall first appear in the shock
upstream then in the downstream and relevant spectral features
shall first appear on the low-frequency band then on the
high-frequency band, inconsistent with the data.

As mentioned, earlier statistical studies on band splits
of both the metric and interplanetary type II bursts have been
presented by \citet{Vrsnak2002,Vrsnak2004}. They also revealed the
almost-constant values of the split frequency ratio. Under the UD
scenario, they deduced the shock Mach number which was found not
to vary much either. Their main purpose was to infer the
magnitudes and radial profiles of the magnetic field strength
(\textbf{B}) from the corona to the interplanetary space through
the following procedure.

First, the shock compression ratio is obtained with the
observed split width. Then, the Mach number is calculated with the
Rankine-Hugoniot jump conditions, assuming a specific plasma
{$\beta$} and a shock geometry (being parallel or perpendicular).
Another step is to get the shock velocity by fitting the spectra
using the prescribed density model. After this, the Alfv{\'{e}}n
speed and further the magnetic field strength can be obtained with
the same density model.
Through this procedure, they deduced reasonable profiles
of \textbf{B}, which are consistent with other diagnostic results
\citep[see][]{Vrsnak2002}. It seems that this may provide a
support to their working assumption that is the UD scenario.
However, we point out that the above process involves many free
parameters and suffers from large uncertainty. For example, the
shock speed is determined by fitting the spectra with the
specified density model. And it is well known that this fitting
process works properly only when the radio source moves outwards
along the density gradient. Also it is a fact that the corona is
highly time-varying and structured and can hardly be described by
a fixed density model.

In addition, since the obtained compression ratio as well
as the deduced Mach number does not change much under the UD
scenario, the radial profile of \textbf{B} is actually controlled
by other factors. The main contribution should be the density
profile, which is used to determine the shock speed by fitting the
dynamic spectra and to deduce \textbf{B} from the obtained
Alfv\'en speed. The reason that the \textbf{B} profile looks
reasonable is mainly because that the adopted density model is
reasonable.

Thus we suggest that the seemingly-reasonable magnitude
and radial profile of B deduced with the UD scenario contains
large uncertainty, the profile is mainly determined by the density
dependence over distance as well as the type-II spectral drift. It
should not be taken as an evidence to support the UD scenario.

In summary, the result of our study, together with that of
\citet{Du2014}, do not favor the UD scenario, which however has
been used in many studies for coronal shock diagnostics, as
mentioned. The exact physical mechanism accounting for the type II
band splits remains unknown. The seemingly constancy of the
frequency split ratio from event to event and during an individual
event tend to suggests that the band splits are given
intrinsically by the emission mechanism and not associated with
outside geometrical profiles, which vary too much to account for
the observation. The split mechanism, whatever it is, shall be
able to confine the frequency split ratio within a relatively
narrow range. Further theoretical and observational studies are
demanded before a definite conclusion can be reached.

\acknowledgements

We are grateful to the RSTN, SOHO/LASCO, GOES and SolarSoft
Latest Events Archive teams for making their data available to us.
We thank Prof. Ching-Sheng Wu for helpful suggestion. This work
was supported by grants NSBRSF 2012CB825601, NNSFC 41274175, 41331068
and the Natural Science Foundation of Shandong Province (ZR2013DQ004
and ZR2014DQ001). Gang Li's work at UAHuntsivlle was supported by NSF
grants ATM-0847719 and AGS1135432.

\begin{table*}[!t]
\centering
\scriptsize
\tabcolsep=2pt
\caption{Some basic parameters of the type-II events. See text for details.}
\renewcommand{\multirowsetup}{\centering}
\begin{tabular}{*{11}{c}}
\hline
\multirow{6}{1.5cm}{Events}&\multirow{6}{1.8cm}{Start-end times of type-IIs}&\multirow{6}{2cm}{Selected interval of study}&\multirow{6}{1.3cm}{Start-end frequencies (lower band, MHz)}&\multirow{6}{1cm}{Df/Dt (MHz /s)}&\multirow{6}{0.9cm}{$\gamma$}&\multirow{6}{1.2cm}{Fitting source speed (km/s)}&\multirow{6}{1.0cm}{CME  linear speed (km/s)}&\multirow{6}{1.2cm}{First C2 appearance time}&\multirow{6}{1.2cm}{Flare (source)}&\multirow{6}{1.5cm}{Flare start \& peak times}\\
&&&&&&&&&&\\
&&&&&&&&&&\\
&&&&&&&&&&\\
&&&&&&&&&&\\
&&&&&&&&&&\\\hline
\hline
\multicolumn{11}{c}{Group A}\\
\hline
20010126&06:05-06:15& 06:09-06:14&64-40&-0.07 &1.19&555&314&06:30&M1(N10E63)$^{*}$&05:50-06:06$^{*}$\\\hline
20020125&02:26-02:45& 02:27-02:38&65-35&-0.05 &1.19&400&213&03:30&C7(N13E10)$^{*}$&02:10-02:26$^{*}$\\\hline
20020511&11:32-11:48& 11:36-11:41&48-36&-0.04 &1.19&429&235&12:26&M1.5(S06W31)$^{*}$&11:16-12:06$^{*}$\\\hline
20021004&22:44-23:00& 22:47-22:56&60-34&-0.04 &1.19&407&310&23:30&M2.9(N11E43)&22:35-22:43\\\hline
20030709&16:34-16:53& 16:42-16:52&58-39&-0.03 &1.23&273&N/A&N/A&C5.8(N13W88)&16:24-16:37\\\hline
20031118&07:47-08:07& 07:49-07:53&80-56&-0.10 &1.16&506&1223&08:06&M3.2(N01E19)&07:23-07:52\\\hline
20100613&05:38-05:58& 05:40-05:43&63-44&-0.13 &1.24&937&320&06:06&M1.0(S24W82)&05:30-05:39\\\hline
20140108&03:48-03:58& 03:50-03:53&73-52&-0.14 &1.22&804&643&04:12&M3.6(N11W88)&03:39-03:47\\\hline
\hline
\multicolumn{11}{c}{Group B}\\
\hline
20040604&07:44-08:00& 07:47-07:48&44-38&-0.14 &1.21&1473&1306&07:50&N/A&N/A\\\hline
20050319&07:07-07:16& 07:10-07:12&52-48&-0.06 &1.22&442&369&07:36&C1.1(S19W52)&06:24-07:06\\\hline
20060430&01:39-02:02& 01:40-01:41&75-56&-0.23 &1.26&1163&428&02:06&C5.4(N16E74)&01:33-01:57\\\hline
20071231&00:53-01:11& 00:54-00:58&83-54&-0.11 &1.28&550&995&01:31&C8.8(S08E81)&00:37-01:11\\\hline
20111117&07:28-07:31& 07:28-07:30&74-55&-0.15 &1.18&826&458&07:48&C6.0(S19E08)&07:16-07:27\\\hline
20120118&23:22-23:32& 23:26-23:29&58-45&-0.03 &1.14&630&270&23:48&C5.1(N25W44)&22:57-23:20\\\hline
20120806&04:44-04:48& 04:44-04:46&61-47&-0.08 &1.15&590&198&05:12&M1.6(S14E88)&04:33-04:38\\\hline
20131119&10:25-10:34& 10:27-10:32&76-48&-0.09 &1.27&533&740&10:36&X1.0(S13W69)&10:14-10:26\\\hline
2014022001&03:22-03:24& 03:23-03:25&52-39&-0.09 &1.18&803&410$^{*}$&04:00$^{*}$&C3.3(S13E36)&03:15-03:35\\\hline
2014022002&03:23-03:26& 03:22-03:24&53-44&-0.08 &1.17&640&410$^{*}$&04:00$^{*}$&C3.3(S13E36)&03:15-03:35\\\hline
2014090701&02:01-02:03& 02:01-02:02&95-77&-0.23 &1.16&735&487&02:24&C7.5(S21E63)&01:53-02:04\\\hline
2014090702&02:03-02:06& 02:03-02:06&75-53&-0.14 &1.17&784&487&02:24&C7.5(S21E63)&01:53-02:04\\\hline
\end{tabular}
\end{table*}

\begin{figure}
\includegraphics[width=0.95\textwidth]{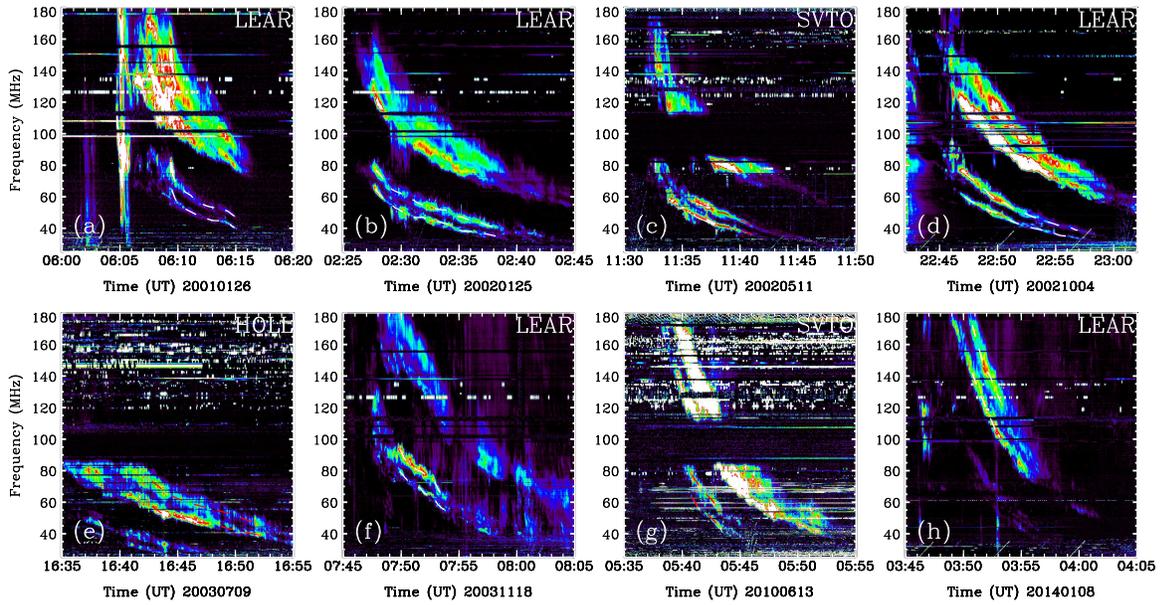}
\caption{The dynamic spectra observed by the RSTN network for
group A events. Dashed lines are backbones given by the
intensity maximum.}\label{Fig1}
\end{figure}

\begin{figure}
\includegraphics[width=0.95\textwidth]{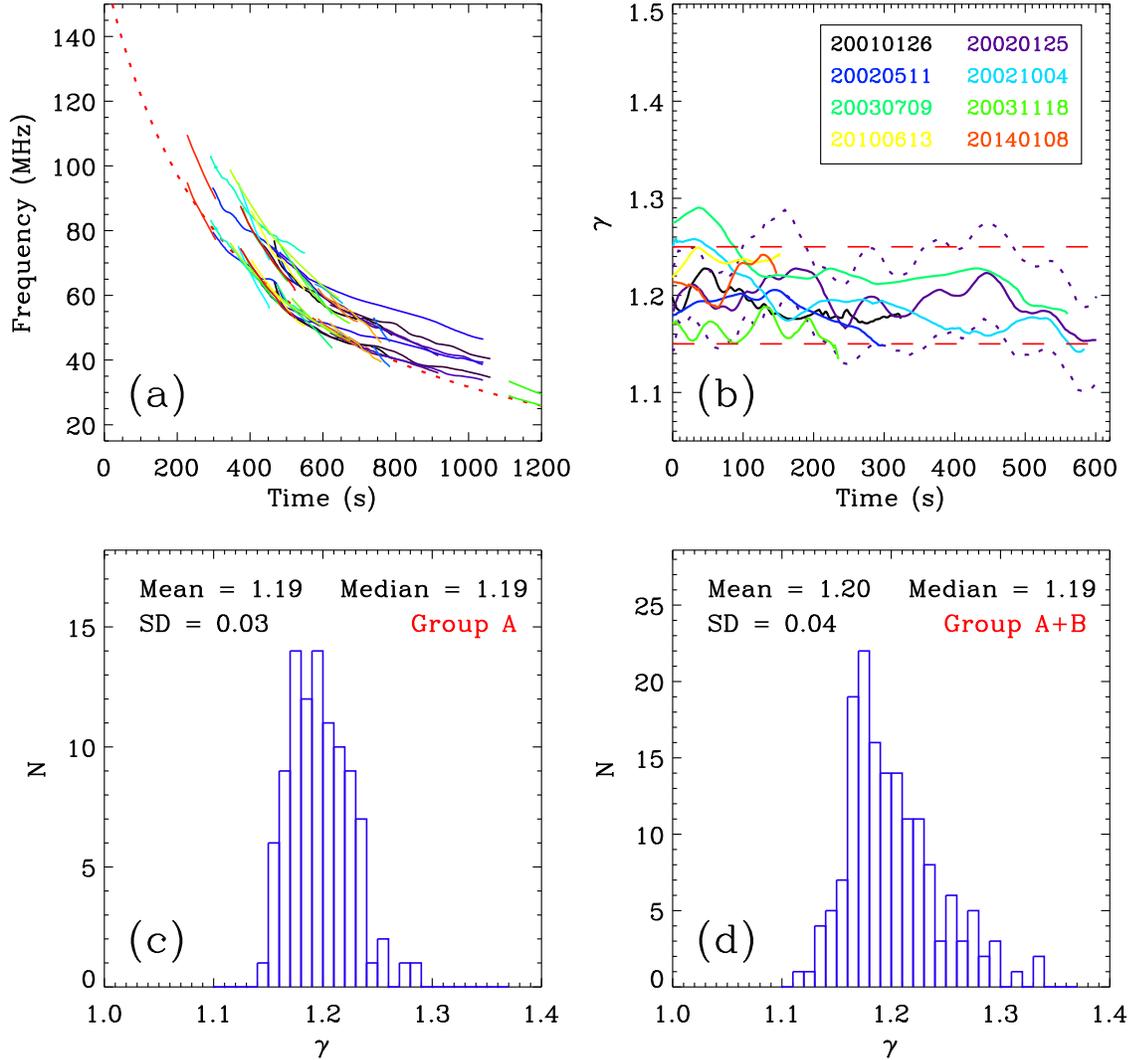}
\caption{(a) Spectral backbones of Group A events plotted together
with the 2X Saito density model (the dashed line). (b) Temporal
evolution of band-split frequency ratios ($\gamma$) with a fixed
starting time. The two dashed lines are measurement uncertainty
for Event 20020125 (see text for details). (c) and (d) are
histograms of the number of data points versus the one-minute
averaged values of {$\gamma$} for Group A and Group A+B events.
}\label{Fig2}
\end{figure}

\begin{figure}
\includegraphics[width=0.95\textwidth]{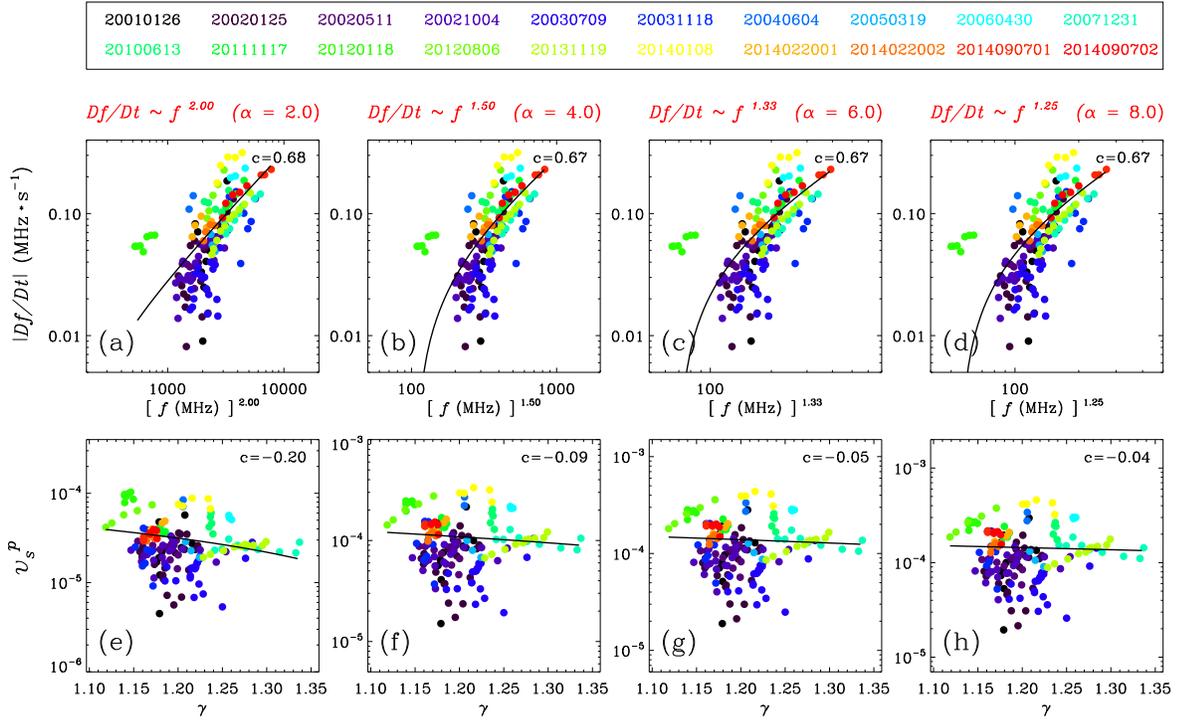}
\caption{(a)-(d) One-minute averaged spectral drift
($\frac{Df}{Dt}$) versus $f^{\frac{\alpha+2}{\alpha}}$ with
{$\alpha$} set to 2, 4, 6, and 10 for all events of our study.
(e)-(h) Deduced shock velocity proxy ($v_{s}^{p}$) versus
band-split frequency ratio $\gamma$ for different values of
$\alpha$ (written on top of each column). Linear fitting lines
(solid black) and corresponding correlation coefficients $c$ are
oresented in each panel.}\label{Fig3}
\end{figure}

\begin{figure}
\includegraphics[width=0.95\textwidth]{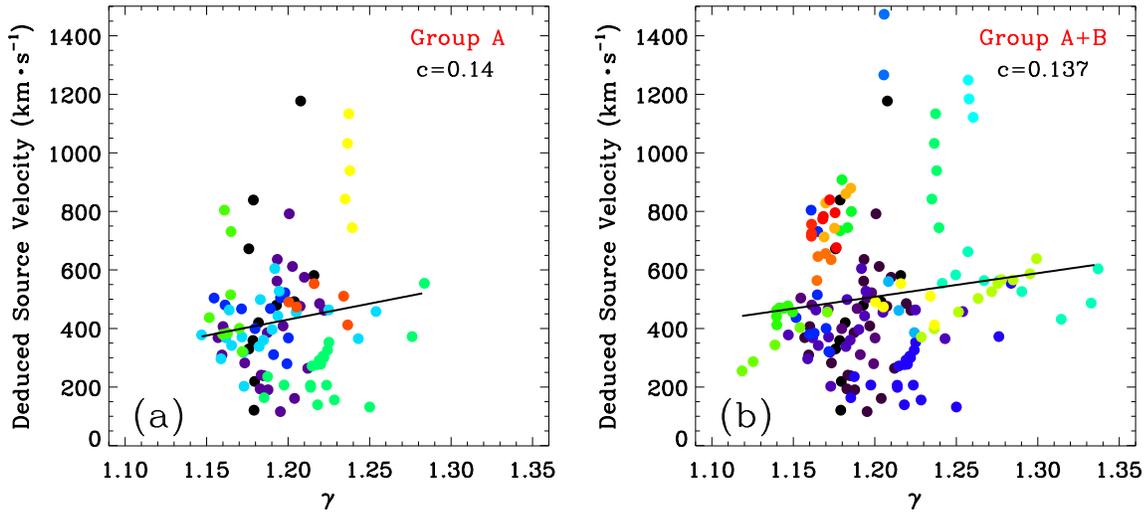}
\caption{ Deduced source velocities versus the band-split
frequency ratio ($\gamma$). The velocities are given by fitting
the ratio spectra of the lower split band with the 2X Saito
density model. }\label{Fig4}
\end{figure}

\end{document}